\documentclass[onecollarge,smallextended,numbook]{svjour2}
\def\makeheadbox{}

\usepackage{color}
\usepackage[hidelinks]{hyperref}
\usepackage{graphicx}
\usepackage{amsfonts}

\def\Eq#1{\label{#1}}
\def\equ#1{(\ref{#1})}

\let\a=\alpha \let\b=\beta          
  \let\h=\eta       \let\l=\lambda
                 \let\p=\pi     
\let\s=\sigma \let\t=\tau        
  \let\ps=\psi        
 \let\D=\Delta

\def\*{\vskip 3mm}
\def\0{\noindent}
\def\be{\begin{equation}}
\def\ee{\end{equation}}
\def\bea{\begin{eqnarray}}
\def\eea{\end{eqnarray}}
\def\nn{\nonumber}

\def\ie{{\it i.e.}\ }

\def\defi{{\buildrel def\over=}}

\def\otto{\,{\kern-1.truept\leftarrow\kern-5.truept\to\kern-1.truept}\,}

\def\tende#1{\,\vtop{\ialign{##\crcr\rightarrowfill\crcr
 \noalign{\kern-1pt\nointerlineskip} \hskip3.pt${\scriptstyle
 #1}$\hskip3.pt\crcr}}\,}
\def\ie{{\it i.e.\ }}

\def\Bt   {{\mbox{\boldmath$ \tau$}}}

\def\Bo   {{\mbox{\boldmath$ \omega$}}}

\let\up\uparrow
\let\down\downarrow
\def\({\left(}
\def\){\right)}

\begin{document}

\title{Kondo effect in the hierarchical $s-d$ model}

\author{Giovanni Gallavotti \and Ian Jauslin}

\institute{
Giovanni Gallavotti \at
INFN-Roma1 and Rutgers University,\\
P.le Aldo Moro 2,\\
00185 Roma, Italy\\
\email{giovanni.gallavotti@roma1.infn.it}\\
homepage {http://ipparco.roma1.infn.it/$\sim$giovanni}
\and
Ian Jauslin \at
University of Rome ``La Sapienza'', Dipartimento di Fisica,\\
P.le Aldo Moro 2,\\
00185 Roma, Italy\\
\email{ian.jauslin@roma1.infn.it}\\
homepage {http://ian.jauslin.org/}
}

\date{December 9, 2015}

\maketitle

\begin{abstract}
The $s-d$ model describes a chain of spin-1/2 electrons interacting magnetically
with a two-level impurity. It was introduced to study the Kondo effect, in which
the magnetic susceptibility of the impurity remains finite in the 0-temperature
limit as long as the interaction of the impurity with the electrons is
anti-ferromagnetic. A variant of this model was introduced by Andrei, which
he proved was exactly solvable via Bethe Ansatz. A hierarchical version
of Andrei's model was studied by Benfatto and the authors. In the present letter, that
discussion is extended to a hierarchical version of the $s-d$ model.
The resulting analysis is very similar to the hierarchical Andrei model,
though the result is slightly simpler.
\end{abstract}

\*

The $s-d$ model was introduced by Anderson \cite{An961} and used by Kondo
\cite{Ko964} to study what would
subsequently be called the {\it Kondo effect}. It describes a chain of
electrons interacting with a fixed spin-1/2 magnetic impurity.
One of the manifestations of the effect is that when the coupling
is anti-ferrmoagnetic, the magnetic susceptibility of the impurity remains
finite in the 0-temperature limit, whereas it diverges for ferromagnetic
and for vanishing interactions.

A modified version of the $s-d$ model was introduced by Andrei \cite{An980},
which was shown to be exactly solvable by Bethe Ansatz. In \cite{BGJ015},
a hierarchical version of Andrei's model was introduced and shown to exhibit
a Kondo effect. In the present letter, we show how the argument can be adapted
to the $s-d$ model.

We will show that in the hierarchical $s-d$ model, the computation
of the susceptibility reduces to iterating an {\it explicit} map relating
6 {\it running coupling constants} (rccs), and that this map can be
obtained by restricting the flow equation for the hierarchical Andrei
model \cite{BGJ015} to one of its invariant manifolds. The physics of
both models are therefore very closely related, as had already been
argued in \cite{BGJ015}. This is particularly noteworthy
since, at 0-field, the flow in the hierarchical Andrei model is relevant,
whereas it is marginal in the hierarchical $s-d$ model, which shows that
the relevant direction carries little to no physical significance.
\*

The $s-d$ model \cite{Ko964} represents a chain of non-interacting spin-1/2
fermions, called {\it electrons}, which interact with an isolated spin-1/2
{\it impurity} located at site 0. The Hilbert space of the system is 
$\mathcal F_L\otimes\mathbb C^2$ in which $\mathcal F_L$ is the Fock space of
a length-$L$ chain of spin-1/2 fermions (the electrons) and $\mathbb C^2$
is the state space for the two-level impurity.
The Hamiltonian, in the presence of a magnetic field of amplitude $h$ in the
direction $\Bo\equiv(\Bo_1,\Bo_2,\Bo_3)$, is
\begin{eqnarray}
H_K&=&H_0+V_0+V_h\defi H_0+V\nn\\
H_0&=&
\sum_{\a\in\{\up,\down\}}\Big(\sum_{x=-{L}/2}^{{L}/2-1}
c^+_\a(x)\,\left(-\frac{\D}2-1\right)\,c^-_\a(x) \Big)\nn\\
V_0&=&-\l_0\sum_{j=1,2,3\atop\a_1,\a_2}
c^+_{\a_1}(0)\s^j_{\a_1,\a_2}c^-_{\a_2}(0)\,
\t^j\Eq{e1}\\
V_h&=& -h \,\sum_{j=1,2,3}\Bo_j \t^j\nn
\end{eqnarray}
where $\l_0$ is the interaction strength, $\Delta$ is the discrete Laplacian
$c_\a^\pm(x),\,\a=\uparrow,\downarrow$ are creation and annihilation
operators acting on {\it electrons}, and $\s^j=\t^j,\,j=1,2,3$, are Pauli
matrices. The operators $\t^j$ act on the {\it impurity}. The boundary
conditions are taken to be periodic.

In the {\it Andrei model} \cite{An980}, the impurity is represented by a fermion instead of a two-level system,
that is the Hilbert space is replaced by $\mathcal F_L\otimes\mathcal F_1$, and the 
Hamiltonian is defined by replacing $\t^j$ in Eq.\equ{e1} by $d^+\t^jd^-$
in which $d_\a^\pm(x),\,\a=\uparrow,\downarrow$ are creation and annihilation
operators acting on the impurity.
\*

The partition function $Z={\rm Tr}\, e^{-\b H_K}$ can be expressed formally as a
functional integral:
\be
Z=\mathrm{Tr}\int P(d\psi)\, \sum_{n=0}^\infty(-1)^n\int_{0<t_1<\cdots<t_n<\beta}\kern-50pt dt_1\cdots dt_n\, \mathcal V(t_1)\cdots\mathcal V(t_n)
\Eq{e2}\ee
in which $\mathcal V(t)$ is obtained from $V$ by replacing $c_\a^\pm(0)$ in
Eq.\equ{e1} by a {\it Grassmann} field $\psi_\a^\pm(0,t)$, $P(d\psi)$ is a
{\it Gaussian Grassmann measure} over the fields $\{\psi_\a^\pm(0,t)\}_{t,\a}$
whose {\it propagator} (\ie {\it covariance}) is, in the $L\to\infty$ limit,
$$
g(t,t')=\frac1{(2\p)^2}\int dk dk_0 \frac{e^{i
k_0(t-t')}}{i k_0-\cos k},
$$
and the trace is over the state-space of the spin-1/2 impurity, that is
a trace over $\mathbb C^2$.
\*

We will consider a {\it hierarchical} version of the
$s-d$ model. The hierarchical model defined below is {\it inspired} by the
$s-d$ model in the same way as the hierarchical model defined in 
\cite{BGJ015} was inspired by the Andrei model. We will not give any
details on the justification of the definition, as such considerations are
entirely analogous to the discussion in \cite{BGJ015}.

The model is defined by introducing a family of {\it hierarchical fields} and
specifying a {\it propagator} for each pair of fields. The average of any
monomial of fields is then computed using the Wick rule.

Assuming $\b=2^{N_\b}$ with $N_\b=\log_2\b\in\mathbb N$, the
time axis $[0,\b)$ is paved with boxes (\ie intervals) of size $2^{-m}$ for
every $m\in\{0,-1,\ldots,-N_\beta\}$: let
\be
{\mathcal Q}_m\defi\left\{[i  2^{|m|}, 
(i+1) 2^{|m|})\right\}_{i=0,1,\cdots
2^{N_\beta-|m|}-1,\atop m=0,-1,\ldots\kern1.4cm}.
\Eq{e3}\ee
Given a box $\Delta\in{\mathcal Q}_m$, let $t_\Delta$ denote the center
of $\Delta$, and given a point $t\in R$, let
$\Delta^{[m]}(t)$ be the (unique) box on scale $m$ that contains $t$.
We further decompose each box $\Delta\in\mathcal Q_m$ into two {\it half boxes}:
for $\h\in\{-,+\}$, let
\be
\Delta_{\h}\defi\Delta^{[m+1]}(t_{\Delta}+\h2^{-m-2})
\Eq{e4}\ee
for $m\le 0$. Thus $\Delta_{-}$ can be called the ``lower half''
of $\Delta$ and $\Delta_{+}$ the ``upper half''.

The elementary fields used to define the hierarchical $s-d$ model will be
{\it constant on each half-box} and will be denoted by
$\ps_\a^{[m]\pm}(\Delta_{\eta})$ for $m\in\{0,-1,\cdots,$ $-N_\beta\}$,
$\Delta\in\mathcal Q_m$, $\eta\in\{-,+\}$,
$\a\in\{\uparrow,\downarrow\}$.

The propagator of the hierarchical $s-d$ model is defined as
\be
\left<\ps_{\a}^{[m]-}(\Delta_{-\h})\ps_{\a}^{[m]+}(\Delta_{\h})\right
>\defi \h
\Eq{e5}\ee
for $m\in\{0,-1,\cdots,$ $-N_\beta\}$,
$\Delta\in\mathcal Q_m$, $\eta\in\{-,+\}$,
$\a\in\{\uparrow,\downarrow\}$. The propagator of any other pair of
fields is set to 0.

Finally, we define
\be
\ps^\pm_\a(t)\defi \sum_{m=0}^{-N_\beta} 2^{\frac{m}2}\ps_\a^{[m]\pm}(\Delta^{[m+1]}(t)).
\Eq{e6}\ee 
The partition function for the hierarchical $s-d$ model is
\be
Z=\mathrm{Tr}\left<
\sum_{n=0}^\infty(-1)^n\int_{0<t_1<\cdots<t_n<\beta}\kern-50pt dt_1\cdots dt_n\, \mathcal V(t_1)\cdots\mathcal V(t_n)
\right>
\Eq{e7}\ee
in which the $\ps^\pm_\a(0,t)$ in $\mathcal V(t)$ have been replaced by
the $\ps_\a^\pm(t)$ defined in Eq.\equ{e6}:
\begin{equation}
\mathcal V(t)\defi-\l_0\sum_{j=1,2,3\atop\a_1,\a_2}
\ps^+_{\a_1}(t)\s^j_{\a_1,\a_2}\ps^-_{\a_2}(t)\,
\t^j
-h \,\sum_{j=1,2,3}\Bo_j \t^j.
\Eq{e8}\end{equation}
This concludes the definition
of the hierarchical $s-d$ model.
\*

We will now show how to compute the partition function Eq.\equ{e7}
using a renormalization group iteration. We first rewrite
\begin{equation}
\sum_{n=0}^\infty(-1)^n\int_{0<t_1<\cdots<t_n<\beta}\kern-50pt dt_1\cdots dt_n\, \mathcal V(t_1)\cdots\mathcal V(t_n)
=\prod_{\Delta\in\mathcal Q_0}\prod_{\eta=\pm}\left(\sum_{n=0}^\infty\frac{(-1)^n}{2^nn!}\mathcal V(t_{\Delta_\eta})^n\right)
\Eq{e9}\end{equation}
and find that
\be
\sum_{n=0}^\infty\frac{(-1)^n}{2^nn!}\mathcal V(t_{\Delta_\eta^{[0]}})^n
=C\left(1+\sum_{p}\ell_p^{[0]}O_{p,\eta}^{[\le 0]}(\Delta^{[0]})\right)
\Eq{e10}\ee
with
\begin{eqnarray}
O_{0,\eta}^{[\le 0]}(\Delta)\defi\frac12\mathbf A^{[\le 0]}_\eta(\Delta)\cdot\Bt,&\quad&
O_{1,\eta}^{[\le 0]}(\Delta)\defi\frac12\mathbf A^{[\le 0]}_\eta(\Delta)^2,\nn\\
O_{4,\eta}^{[\le 0]}(\Delta)\defi\frac12\mathbf A^{[\le 0]}_\eta(\Delta)\cdot\Bo,&\quad&
O_{5,\eta}^{[\le 0]}(\Delta)\defi\frac12 \Bt\cdot\Bo,\Eq{e11}\\
O_{6,\eta}^{[\le 0]}(\Delta)\defi\frac12(\mathbf A^{[\le 0]}_\eta(\Delta)\cdot\Bo)(\Bt\cdot\Bo),&\quad&
O_{7,\eta}^{[\le 0]}(\Delta)\defi\frac12(\mathbf A^{[\le 0]}_\eta(\Delta)^2)(\Bt\cdot\Bo)\nn
\end{eqnarray}
(the numbering is meant to recall that in \cite{BGJ015}) in which $\Bt=(\t^1,\t^2,\t^3)$ and $\mathbf A_\eta^{[\le 0]}(\Delta)$ is a vector of polynomials in the
fields whose $j$-th component for $j\in\{1,2,3\}$ is
\be
A_\eta^{[\le 0]j}(\Delta)\defi\sum_{(\alpha,\alpha')\in\{\uparrow,\downarrow\}^2}
\ps_\a^{[\le 0]+}(\Delta_\eta)\sigma^j_{\a,\a'}\ps_{\a'}^{[\le 0]-}(\Delta_\eta)
\Eq{e12}\ee
$\ps_\a^{[\le 0]\pm}:=\sum_{m\le0}2^{\frac m2}\ps_\a^{[m]\pm}$, and
\begin{eqnarray}
C&=&\cosh(\tilde h),\quad
\ell_0^{[0]}=\frac1C\frac{\lambda_0}{\tilde h}\sinh(\tilde h)\nn\\
\ell_1^{[0]}&=&\frac1C\frac{\lambda_0^2}{12\tilde h}(\tilde h\cosh(\tilde h)+2\sinh(\tilde h))\nn\\
\ell_4^{[0]}&=&\frac1C\lambda_0\sinh(\tilde h),\quad
\ell_5^{[0]}=\frac2C\sinh(\tilde h)\Eq{e13}\\
\ell_6^{[0]}&=&\frac1C\frac{\lambda_0}{\tilde h}(\tilde h\cosh(\tilde h)-\sinh(\tilde h))\nn\\
\ell_7^{[0]}&=&\frac1C\frac{\lambda_0^2}{12\tilde h^2}(\tilde h^2\sinh(\tilde h)+2\tilde h\cosh(\tilde h)-2\sinh(\tilde h))\nn
\end{eqnarray}
in which $\tilde h:=h/2$.

By a straightforward induction, we find that the partition function
Eq.\equ{e7} can be computed by defining
\be
C^{[m]}\mathcal W^{[m-1]}(\Delta^{[m]})\defi\left<\prod_\eta\left(\mathcal W^{[m]}(\Delta^{[m]}_\eta)\right)\right>_m
\Eq{e14}\ee
in which $\left<\cdot\right>_m$ denotes the average over $\psi^{[m]}$, $C^{[m]}>0$ and
\be
\mathcal W^{[m-1]}(\Delta^{[m]})=1+\sum_p\ell_p^{[m]}O_p^{[\le m]}(\Delta^{[m]})
\Eq{e15}\ee
in terms of which
\be
Z=C^{2|\mathcal Q_0|}\prod_{m=-N(\beta)+1}^0(C^{[m]})^{|\mathcal Q_{m-1}|}
\Eq{e16}\ee
in which $|\mathcal Q_m|=2^{N(\beta)-|m|}$ is the cardinality of $\mathcal Q_m$.
In addition, similarly to \cite{BGJ015}, the map relating $\ell_p^{[m]}$
to $\ell_p^{[m-1]}$ and $C^{[m]}$ can be computed explicitly from Eq.\equ{e14}:
\begin{eqnarray}
C^{[m]} &=& 1
+\frac{3}{2}\ell_{0}^2
+\ell_{0}\ell_{6}
+9\ell_{1}^2
+\frac{\ell_{4}^2}{2}
+\frac{\ell_{5}^2}{4}
+\frac{\ell_{6}^2}{2}
+9\ell_{7}^2
\nn\\
\ell^{[m-1]}_{0} &=& \frac1C\Big(\ell_{0}
-\ell_{0}^2
+3\ell_{0}\ell_{1}
-\ell_{0}\ell_{6}
\Big)\nn\\
\ell^{[m-1]}_{1} &=& \frac1C\Big(\frac{\ell_{1}}{2}
+\frac{\ell_{0}^2}{8}
+\frac{\ell_{0}\ell_{6}}{12}
+\frac{\ell_{4}^2}{24}
+\frac{\ell_{5}\ell_{7}}{4}
+\frac{\ell_{6}^2}{24}
\Big)\nn\\
\ell^{[m-1]}_{4} &=& \frac1C\Big(\ell_{4}
+\frac{\ell_{0}\ell_{5}}{2}
+3\ell_{0}\ell_{7}
+3\ell_{1}\ell_{4}
+\frac{\ell_{5}\ell_{6}}{2}
+3\ell_{6}\ell_{7}
\Big)\nn\\
\ell^{[m-1]}_{5} &=& \frac1C\Big(2\ell_{5}
+2\ell_{0}\ell_{4}
+36\ell_{1}\ell_{7}
+2\ell_{4}\ell_{6}
\Big)\Eq{e17}\\
\ell^{[m-1]}_{6} &=& \frac1C\Big(\ell_{6}
+\ell_{0}\ell_{6}
+3\ell_{1}\ell_{6}
+\frac{\ell_{4}\ell_{5}}{2}
+3\ell_{4}\ell_{7}
\Big)\nn\\
\ell^{[m-1]}_{7} &=& \frac1C\Big(\frac{\ell_{7}}{2}
+\frac{\ell_{0}\ell_{4}}{12}
+\frac{\ell_{1}\ell_{5}}{4}
+\frac{\ell_{4}\ell_{6}}{12}
\Big)\nn
\end{eqnarray}
in which the $^{[m]}$ have been dropped from the right hand side.
\*

The flow equation Eq.\equ{e17} can be recovered from that of the
hierarchical Andrei model studied in \cite{BGJ015} (see in particular
\cite[Eq.(C1)]{BGJ015}) by restricting
the flow to the invariant submanifold defined by
\be
\ell_2^{[m]}=\frac13,\quad
\ell_3^{[m]}=\frac16\ell_1^{[m]},\quad
\ell_8^{[m]}=\frac16\ell_4^{[m]}.
\Eq{e18}\ee
This is of particular interest since $\ell_2^{[m]}$ is a relevant coupling and
the fact that it plays no role in the $s-d$ model indicates that it has little
to no physical relevance.

The qualitative behavior of the flow is therefore the same as that described
in \cite{BGJ015} for the hierarchical Andrei model. In particular
the susceptibility, which can be computed by deriving $-\beta^{-1}\log Z$ with respect
to $h$, remains finite in the 0-temperature limit as long as $\lambda_0<0$,
that is as long as the interaction is anti-ferromagnetic.

\begin{acknowledgements}
We are grateful to G.~Benfatto for many enlightening discussions on the $s-d$ and Andrei's models.
\end{acknowledgements}

\bibliographystyle{spmpsci} 

\begin{thebibliography}{1}
\providecommand{\url}[1]{#1}
\providecommand{\urlprefix}{URL }
\expandafter\ifx\csname urlstyle\endcsname\relax
  \providecommand{\doi}[1]{DOI~\discretionary{}{}{}#1}\else
  \providecommand{\doi}{DOI~\discretionary{}{}{}\begingroup
  \urlstyle{rm}\Url}\fi

\bibitem{An961}
Anderson, P.: {Local magnetized states in metals}.
\newblock Physical Review \textbf{124}, 41--53 (1961)

\bibitem{An980}
Andrei, N.: {Diagonalization of the Kondo Hamiltonian}.
\newblock {Physical Review Letters} \textbf{45}, 379--382 (1980)

\bibitem{BGJ015}
Benfatto, G., Gallavotti, G., Jauslin, I.: Kondo effect in a fermionic
  hierarchical model.
\newblock arXiv: 1506.04381  (2015)

\bibitem{Ko964}
Kondo, J.: {Resistance Minimum in Dilute Magnetic Alloys}.
\newblock Progress of Theoretical Physics \textbf{32}, 37--49 (1964)

\end{thebibliography}
\input\jobname.bbl
\end{document}